\documentclass[12pt]{article}
\usepackage{amsmath}
\usepackage{amssymb}
\usepackage{graphicx}% Include figure files
\usepackage{bm}% bold math
\newcommand{\bftau}{\mbox{\boldmath $\tau$}}

\newcommand{\bfsigma}{\mbox{\boldmath $\sigma$}}
\begin{document}
\title{\bf
 EXCITATIONS OF THE UNSTABLE\\
 NUCLEI $^{48}$Ni AND $^{49}$Ni}
\author{%
S. P. Kamerdzhiev \\
{\it \small Institute of Physics and Power Engeneering,
 249020, Obninsk, Russia}\\$\vphantom{,}$\\
V. I. Tselyaev \\
{\it \small Nuclear Physics Department,
 V. A. Fock Institute of Physics},\\
{\it \small St. Petersburg State University, 198504,
 St. Petersburg, Russia}}
\date{}
\maketitle
\begin{abstract}
The isoscalar E1 and E2 resonances in the proton-rich nuclei
$^{48,49}$Ni and the $\{ f_{7/2} \otimes 3^- \}$ multiplet
in $^{49}$Ni have been calculated  taking into account
the single-particle continuum exactly. The analogous 
calculations for the mirror nuclei $^{48}$Ca and $^{49}$Sc
are presented. The models used are the continuum RPA
for $^{48}$Ni, $^{48}$Ca and the Odd RPA
for $^{49}$Ni, $^{49}$Sc,
the latter has been  developed recently and describes
both single-particle and collective excitations of
an odd nucleus on a common basis.
In all four nuclei we obtained a distinct splitting of the
isoscalar E1 resonance into
$1 \hbar \omega$ and $3 \hbar \omega$ peaks
at about 11 MeV and 30 MeV, respectively. 
The main part of the isoscalar E1 energy-weighted sum rule (EWSR)
is exhausted by the $3 \hbar \omega$ resonances.
The $1 \hbar \omega$ resonances exhaust about 35~\% of this EWSR
in $^{48,49}$Ni and about 22~\% in $^{48}$Ca and $^{49}$Sc.
All seven $\{ f_{7/2} \otimes 3^- \}$ multiplet members in
$^{49}$Ni are calculated to be in the (6--8) MeV
energy region and have noticeable escape widths.
\end{abstract}

\vspace{2em}
\newpage

The main feature of the proton-rich unstable nuclei $^{48}$Ni
and $^{49}$Ni, which have been discovered recently
\cite{Bl1,Bl2}, is that they have a proton separation energy
near zero.
It seems to be of great interest to calculate excitations
of these nuclei, especially in the framework of the same
calculational scheme, which would take into account 
the single-particle continuum reliably,
because its role is obviously
very important in this case, though the corresponding effects
may be not so strong as in the neutron-rich nuclei with the
neutron separation energy near zero.
The comparison of the theoretical predictions with
future measurements of the excitations of these nuclei 
will make  it possible  to obtain an important
information about the nucleon-nucleon effective interaction
in drip-line nuclei and about features of such nuclei,
which are also of astrophysical interest.

    For the doubly-magic $^{48}$Ni, one of the
proper microscopic methods is the theory of finite
Fermi-systems (TFFS) \cite{Mig} with exact accounting for
the single-particle continuum (continuum TFFS). It corresponds
to the continuum random phase approximation (RPA)
making use of the phenomenological Landau-Migdal
particle-hole (p-h) interaction and a single-particle
Woods-Saxon scheme. The parameters describing this interaction
and the single-particle scheme are known 
\cite{ip1,ip2,ch}, at least for stable nuclei,
and it is of great interest to check them for unstable ones. 

For doubly magic nuclei plus or minus one nucleon,
a theory has been developed recently
\cite{KLT}, which was termed the Odd RPA (ORPA).
It corresponds to the continuum RPA for even-even nuclei and,
in addition, consistently 
accounts for properties of odd nuclei.
The ORPA allows a consistent description of  both
the single-particle and collective part of the excitation spectrum
of an odd nucleus without pairing, including giant resonances
in the continuum and splitting of particle (hole)~$\otimes$~phonon
multiplets, on a common basis. The description of these
multiplets  taking the single-particle continuum
into account is especially important and interesting for nuclei
such as  $^{49}$Ni  with the nucleon separation energy near zero
and relatively large  phonon energies.
Other questions concerning the excitations of this nucleus
exist. For example, it was shown
in \cite{KLT} that the isoscalar (IS) E1 strength
in $^{17}$O below 12.5 MeV is completely determined
by the odd nucleon contribution. The question
about a similar effect in $^{49}$Ni arises.

Thus, we have a reasonable and very interesting possibility
to calculate excitations of both nuclei within the same
calculational scheme based on the continuum TFFS as well as
to investigate primarily the role of the small proton separation
energy  in the proton-rich $^{48,49}$Ni nuclei and of the 
single-particle continuum.
This is the aim of the present article.
We will present the results of the
calculations of the IS E1 and E2
giant resonances in $^{48}$Ni and $^{49}$Ni. For $^{49}$Ni
the characteristics of the $\{ f_{7/2} \otimes 3^- \}$
multiplet will be  presented for all 7 resonances.
For comparison and reliability of the results, the
resonances in the mirror nuclei $^{48}$Ca and $^{49}$Sc
and the multiplet in $^{49}$Sc are also calculated.
In the calculations we use the TFFS and Woods-Saxon
potential parameters which are mainly known.
However, in order to have reliable results for the
exotic nuclei, we fit some of the parameters of
the Landau-Migdal interaction and Woods-Saxon  potential
using Hartree-Fock single-particle levels 
obtained with the known Skyrme interaction and
experimental data for $^{48}$Ca nucleus (see below).

In the TFFS approach \cite{Mig} the effective interaction
and the single-particle scheme are described by two sets
of phenomenological parameters and are not connected with
each other. In our calculations the single-particle
potential was taken in the form
\begin{equation}
 U_q({\bf r}) = U^0_q f_{WS}(r)
 + U^{so}_q \frac{\kappa}{r} \frac{d f_{WS}(r)}{dr}
 ({\bf l} \cdot {\bf {\bfsigma}}) + U_C(r)\,,
\label{uws}
\end{equation}
where $f_{WS}(r) = 1/[1 + \exp ((r-R)/a)]$, index $q = n, p$
denotes the sort of the nucleon. The Coulomb potential $U_C$
for protons is that of a uniformly charged sphere.
The parameters $U^{so}_q$, $\kappa$, $R$, and $a$ used here 
are well known \cite{ch}:
$$
U^{so}_p = (Z V_{pp} + N V_{pn})/A\, , \qquad
U^{so}_n = (N V_{nn} + Z V_{np})/A\, ,
$$
$$
V_{pp} = V_{nn} = 19.7\: \mbox{MeV}\, , \qquad
V_{pn} = V_{np} = 87.0\: \mbox{MeV}\, ,
$$
$$
\kappa = 0.263 \left( 1 + 2 (N-Z)/A \right)\: \mbox{fm}^2\,,
$$
$$
R = 1.24\, A^{1/3}\: \mbox{fm}\,, \qquad
a = 0.63\: \mbox{fm}\,.
$$

     The choice of the parameters $U^0_q$ in Eq.~(\ref{uws})
is a more complicated question for the $^{48,49}$Ni nuclei.
The point is that the value of $U^0_p$ cannot be determined
from the standard set of parameters \cite{ch},
which corresponds to $U^0_q = U^{so}_q$ in Eq.~(\ref{uws}), because
in this case it leads to a positive  energy of the last occupied
proton $1f_{7/2}$ level in $^{48}$Ni. To overcome this
difficulty we used the following method. We have calculated the
energies of single-particle proton and neutron $1f_{7/2}$ levels
in $^{48}$Ni in the framework of the self-consistent Hartree-Fock
method with the Skyrme SIII interaction of Ref.~\cite{siii},
which is often used in the calculations of properties of both
stable and unstable nuclei. Then the values of the potential
depths $U^0_p$ and $U^0_n$ in Eq.~(\ref{uws}) were fitted so
that to reproduce the Hartree-Fock values of
$\varepsilon_p (1f_{7/2}) = -  0.86$ MeV and
$\varepsilon_n (1f_{7/2}) = - 16.98$ MeV.
From this input we obtained
$U^0_p = - 52.03$ MeV, $U^0_n = - 60.98$ MeV for
$^{48}$Ni. The same procedure was performed for the $^{48}$Ca
nucleus. In this case the values $U^0_p = - 58.32$ MeV and
$U^0_n = - 49.09$ MeV yield the Hartree-Fock
values of the single-particle energies
$\varepsilon_p (1f_{7/2}) = - 10.16$ MeV and
$\varepsilon_n (1f_{7/2}) = -  9.99$ MeV.
The corresponding experimental values for $^{48}$Ca are
$- 9.63$ MeV and $- 9.95$ MeV,
i.~e. the agreement with experiment is fairly good.

     The Landau-Migdal p-h interaction used in our
calculation is defined by the following ansatz
\begin{equation}
{\cal F} ({\bf r}_1, {\bf r}_2) = C_0
\left( f({\bf r}_1)
+ f'({\bf r}_1) ({\bf {\bftau}}_1 \cdot {\bf {\bftau}}_2)
+ g ({\bf {\bfsigma}}_1 \cdot {\bf {\bfsigma}}_2)
+ g'({\bf {\bftau}}_1 \cdot {\bf {\bftau}}_2)
    ({\bf {\bfsigma}}_1 \cdot {\bf {\bfsigma}}_2) \right)
\delta ({\bf r}_1 - {\bf r}_2)\,,
\label{phint}
\end{equation}
where
\begin{equation}
f({\bf r}) =
f_{ex}+(f_{in}-f_{ex}) \rho ({\bf r}) / \rho (0)\,,
\qquad
f'({\bf r}) =
f'_{ex}+(f'_{in}-f'_{ex}) \rho ({\bf r}) / \rho (0)\,.
\label{ff}
\end{equation}
The total nucleon density $\rho ({\bf r})$ entering
Eqs.~(\ref{ff}) is defined in our approach by the formula
\begin{equation}
\rho ({\bf r}) = \sum_{\lambda , \sigma} n_{\lambda}
|\varphi_{\lambda} ({\bf r}, \sigma)|^2\,,
\label{rho}
\end{equation}
where $n_{\lambda}$ are the occupation numbers,
$\varphi_{\lambda}$ are the single-particle wave functions
calculated within the above-described Woods-Saxon scheme.

     The parameters entering Eq.~(\ref{phint}) are known
\cite{ip1,ip2}, as a rule,
except for one of  the six parameters,
namely the parameter $f_{ex}$, which is fitted very often
to the experimental energies of low-lying phonons, see,
for example, Refs.~\cite{KLT,EPAN}.
In our case this is especially
necessary because, strictly speaking, there is no guarantee
that the Landau-Migdal interaction is suitable for the
drip-line nuclei. Since the  excitations
of $^{48}$Ni are unknown, it is reasonable
to use known excitations of the mirror nucleus $^{48}$Ca
for adjusting $f_{ex}$.
So we have fitted the parameter $f_{ex}$ 
to reproduce (within the continuum TFFS)
the energy of the $3_1^-$ level in $^{48}$Ca at 4.51 MeV.
The result was $f_{ex} = -0.94$. Thus, in all our
calculations the following set of the Landau-Migdal
interaction parameters has been used (the rest of them
being taken from Refs.~\cite{ip1,ip2,EPAN,KLT})
$$
 f_{ex} = - 0.94\,,\quad f_{in} = - 0.002\,,\quad
 f'_{ex} = 2.30\,,\quad f'_{in} = 0.76\,,
$$
$$
 g = 0.05\,,\quad g' = 0.96\,,\quad C_0 = 300\:
 \mbox{MeV}\cdot\mbox{fm}^3\,.
$$

With these parameters we obtained  a reasonable agreement
with experiment for $2_1^+$ level ($E_{th}(2_1^+)$ = 4.15 MeV,
$E_{exp}(2_1^+)$ = 3.83 MeV) and very good agreement
for the $B(E3)\uparrow$ = $B(E3,\, g.s. \to 3_1^-)$ value
($B_{th}(E3)\uparrow$ = 6500 $e^2 fm^6$ and
$B_{exp}(E3)\uparrow$ = 6700 $e^2 fm^6$) in $^{48}$Ca.
The latter is important for the realistic description
of the $\{ f_{7/2} \otimes 3^- \}$ multiplet.

     The calculations of the IS E1 and E2 resonances
in $^{48}$Ni and $^{48}$Ca were performed within continuum TFFS.
For the IS E1 resonance  the "forced consistency"
and  the spurious state suppression procedure was applied.
This procedure and other details of the calculational scheme
have been  described in Refs.~\cite{KLT,kllt}.
The same resonances and the $\{ f_{7/2} \otimes 3^- \}$
multiplets in $^{49}$Ni and $^{49}$Sc have been
calculated within the ORPA model developed in \cite{KLT}.
In the ORPA calculations, only one (entrance) channel was
incorporated and  same interaction parameters and the
single-particle schemes were used as in our TFFS ones.
All the calculations were performed using
the coordinate representation technique in the framework
of the Green function method \cite{sb} which enables
one to take into account the single-particle continuum
exactly. The smearing parameters
used were $\Delta$ = 300 keV for the giant resonances
and $\Delta$ = 0.2 keV for the multiplets.

     One-body transition operators (i.~e. the external field
operators $V^0$ in terms of TFFS)
were taken in the following form:
\begin{equation}
V^{0(E1)}_q = (r^3 - R^2_{IS} r) Y_{1\mu}(\hat{r})
\label{v0e1}
\end{equation}
for isoscalar E1 excitations,
\begin{equation}
V^{0(E2)}_q = r^2 Y_{2\mu}(\hat{r})
\label{v0e2}
\end{equation}
for isoscalar E2 excitations, and
\begin{equation}
V^{0(E3)}_q = e_q r^3 Y_{3\mu}(\hat{r})
\label{v0e3}
\end{equation}
for electromagnetic E3 excitations, where $e_q$ are the nucleon
effective charges in the center-of-mass reference frame
(see, e. g., Ref.~\cite{sb}):
\begin{equation}
e_p = \left[ \left( 1 - \frac{1}{A} \right)^L
+ (Z-1) \left( - \frac{1}{A} \right)^L \right] e\,,
\qquad e_n = Z \left( - \frac{1}{A} \right)^L e
\label{effch}
\end{equation}
with $L=3$. Parameter $R_{IS}$ in Eq.~(\ref{v0e1}) is determined
within the "forced consistency" scheme (see \cite{KLT}).
In the present calculations the numerical values of $R_{IS}$
for all nuclei under consideration are very close (within 1~\%)
to those of $(\frac{5}{3} <r^2>)^{1/2}$
(in this context, see Ref.~\cite{NVGS}).

\begin{figure}
\includegraphics*[scale=0.84]{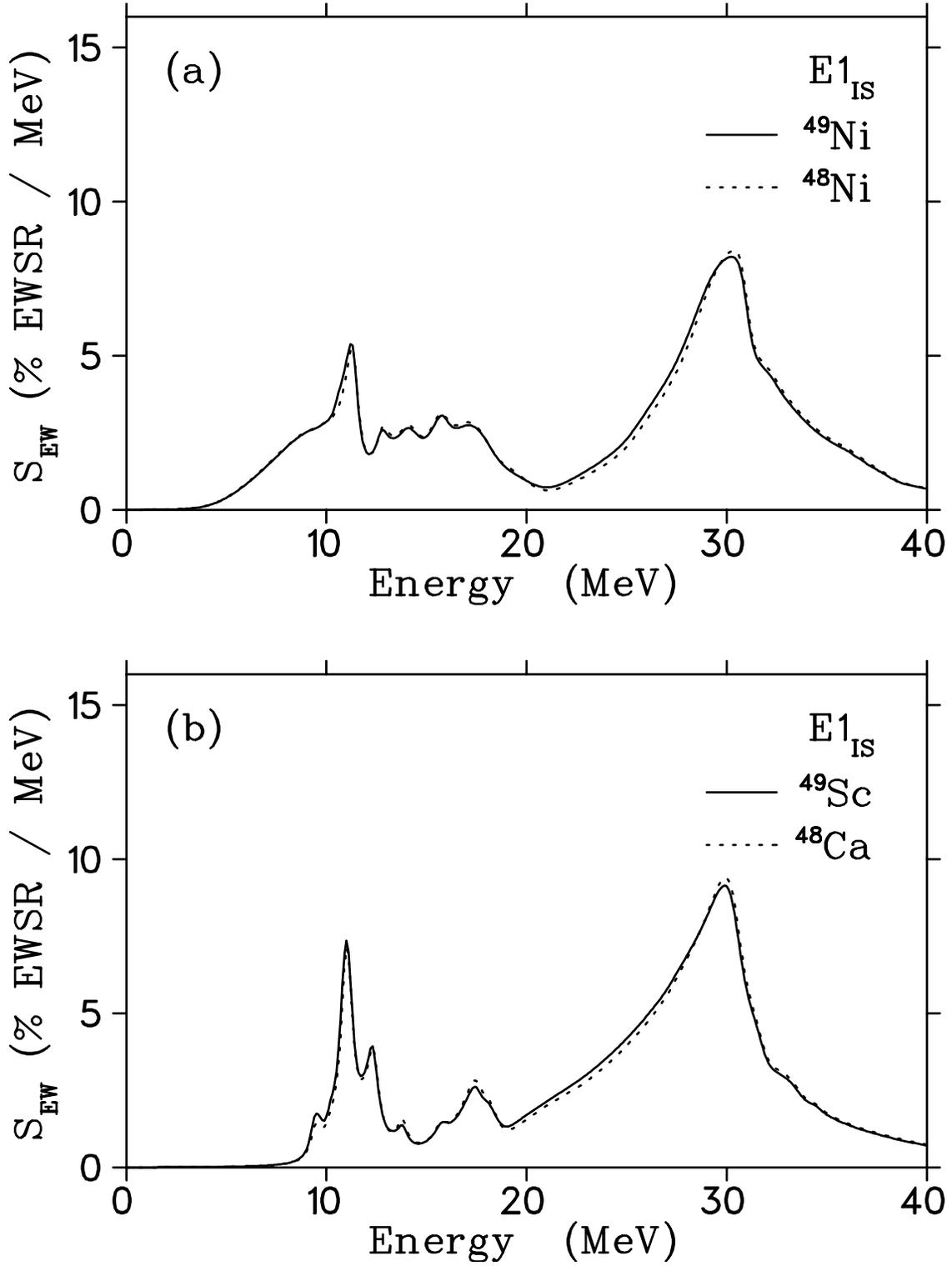}
\caption{\label{fig1}
The normalized energy-weighted strength functions
$S_{EW}(E)$ for the isoscalar (IS) E1 excitations
in $^{49}$Ni (a) and $^{49}$Sc (b) calculated
in the ORPA (solid lines), and the same functions for
$^{48}$Ni (a) and $^{48}$Ca (b)
calculated in the continuum TFFS (dotted lines).
The smearing parameter is  $\Delta$ = 300 keV.}
\end{figure}

\begin{figure}
\includegraphics*[scale=0.84]{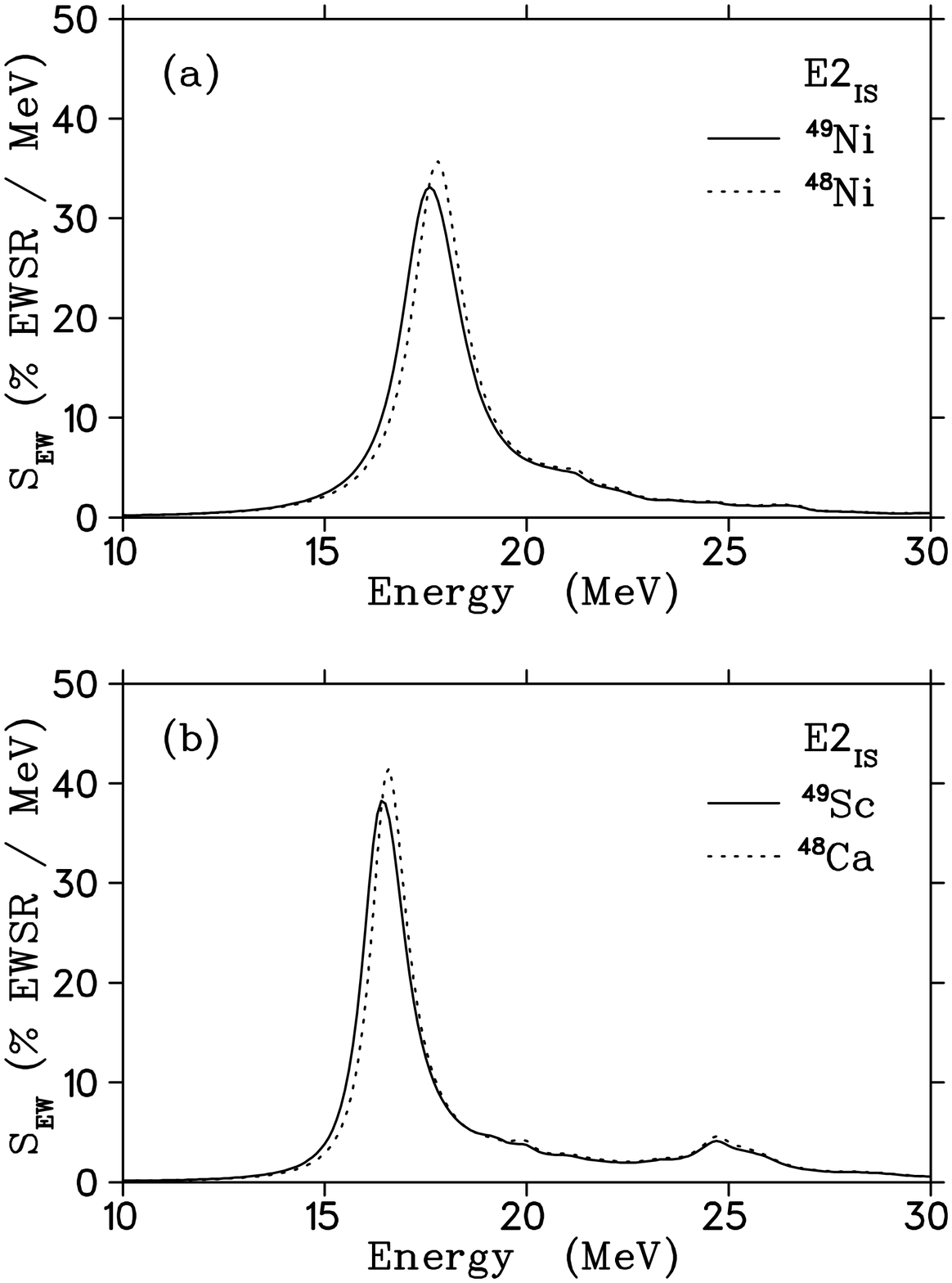}
\caption{\label{fig2}
Same as Fig.~1, but for the isoscalar E2 resonance.}
\end{figure}

     The results for the IS E1 and E2 resonances are presented in
Figs.~\ref{fig1} and ~\ref{fig2} in the form of the normalized
energy-weighted strength functions $S_{EW} (E)$:
\begin{equation}
S_{EW} (E) = \frac{100}{m^{SR}_1} E S(E)
\label{sew}
\end{equation}
where $S(E)$ is the RPA (ORPA) strength function, $m^{SR}_1$
is the corresponding energy-weighted sum rule (EWSR), which
is defined as
\begin{equation}
m^{SR}_1 = \frac{3 \hbar ^2 A}{8 m \pi}
\left( 11 <r^4> - \frac{25}{3} <r^2>^2 \right)
\label{sre1}
\end{equation}
for the IS E1 operator (\ref{v0e1}), and
\begin{equation}
m^{SR}_1 = \frac{25 \hbar ^2 A}{4 m \pi} <r^2>
\label{sre2}
\end{equation}
for the IS E2 operator (\ref{v0e2}). Note that these formulas
differ from the often used definitions of the IS EL EWSR
(see Refs. \cite{NVGS,LB}) by the factor $(2L+1)$
arising due to the summation over the quantum numbers of the
excited states and of an external field in the strength functions.

   The functions $S_{EW}(E)$ for IS E1 excitations
of all the nuclei considered are shown in Fig.~\ref{fig1}.
The IS E1 strength is distributed around two main maxima at
about 11 MeV and 30 MeV, which correspond to the
$1 \hbar \omega$ and $3 \hbar \omega$ IS E1 resonances.
The calculated position of the $3 \hbar \omega$ resonances
is in agreement with the experimental formula
$E_x \sim 114\, A^{-1/3}$ = 31.4 MeV for $A=48$, which was obtained
in Ref.~\cite{CLY} for the upper components of the IS E1 resonances
in $^{90}$Zr, $^{116}$Sn, and $^{208}$Pb.
To estimate the contribution of these resonances
into the IS E1 EWSR (\ref{sre1})
the following integrals were calculated
 \begin{equation}
p (E_{min}; E_{max}) = \int\limits_{E_{min}}^{E_{max}}
S_{EW} (E) d E\,.
 \label{pewsr}
 \end{equation}
The contribution of the $1 \hbar \omega$ resonance
($p(1 \hbar \omega)$)
was defined as $p (0; 21)$ in $^{48}$Ni, $^{49}$Ni
(hereinafter the values of $E_{min}$ and $E_{max}$
are given in MeV), and as $p (0; 19)$ in
$^{48}$Ca, $^{49}$Sc, where the values of 21 MeV and 19 MeV
are the energies of minima separating $1 \hbar \omega$ and
$3 \hbar \omega$ resonances in these nuclei. For the
$3 \hbar \omega$ resonance the contribution
($p(3 \hbar \omega)$)
was defined as $p (21; 40)$ in $^{48}$Ni, $^{49}$Ni,
and as $p (19; 40)$ in $^{48}$Ca, $^{49}$Sc.
The results, which are listed in Table~\ref{tab1},
show that the main part of the IS E1 EWSR is exhausted by the
$3 \hbar \omega$ resonances in all nuclei under consideration.
\begin{table}
\begin{center}
\caption{\label{tab1}
The percentages of the IS E1 EWSR for the
$1 \hbar \omega$ and $3 \hbar \omega$ IS E1 resonances
(see text for details).}
\vspace{1em}
\begin{tabular}{lrrrr}
\hline
\hline
\multicolumn{1}{c}{} &
\multicolumn{1}{c}{$^{48}$Ni} &
\multicolumn{1}{c}{$^{49}$Ni} &
\multicolumn{1}{c}{$^{48}$Ca} &
\multicolumn{1}{c}{$^{49}$Sc}\\
\hline
 $p(1 \hbar \omega)$, \% EWSR & 35 & 35 & 21 & 22 \\
 $p(3 \hbar \omega)$, \% EWSR & 59 & 60 & 73 & 73 \\
\hline
\hline
\end{tabular}
\end{center}
\end{table}
The $1 \hbar \omega$ resonance
exhausts about 35~\% of the IS E1 EWSR in
$^{48,49}$Ni and about 22~\% in $^{48}$Ca and $^{49}$Sc. 

Further, comparing
the results for the doubly-magic and neighbouring odd nuclei
in Fig.~\ref{fig1}, we see that the role of the odd nucleon is
rather small in our case. The  difference from the
IS E1 resonances in $^{16}$O and $^{17}$O (see Ref.~\cite{KLT})
is that in $^{49}$Ni and $^{49}$Sc
the single-particle resonances are
embedded into the region of the core p-h excitations and are only
slightly distinguishable  against their background .

The IS E2 resonances in $^{48}$Ni, $^{49}$Ni,
$^{48}$Ca, and $^{49}$Sc are shown in Fig.~\ref{fig2}.
Their characteristics are given in Table~\ref{tab2}.
\begin{table}
\begin{center}
\caption{\label{tab2}
Energy of the maximum $E_0$, the width at half maximum $\Gamma_0$,
and the percentage of the IS E2 EWSR in the energy interval
10--30 MeV for IS GQR.}
\vspace{1em}
\begin{tabular}{lrrrr}
\hline
\hline
\multicolumn{1}{c}{} &
\multicolumn{1}{c}{$^{48}$Ni} &
\multicolumn{1}{c}{$^{49}$Ni} &
\multicolumn{1}{c}{$^{48}$Ca} &
\multicolumn{1}{c}{$^{49}$Sc}\\
\hline
 $E_0$, MeV      & 17.8 & 17.6 & 16.6 & 16.4 \\
 $\Gamma_0$, MeV & 1.5  & 1.8  & 1.1  & 1.3  \\
 $p(\mbox{IS GQR})$, \% EWSR
                 & 94   & 94   & 91   & 92   \\
\hline
\hline
\end{tabular}
\end{center}
\end{table}
We obtained a reasonable depletion of
the IS E2 EWSR within the 10--30 MeV interval taken.
The positions of the maxima of the calculated strength
functions are in a reasonable agreement with an empirical
formula for the mean energy of the IS giant quadrupole
resonance (GQR)
$\bar E _{IS GQR} \sim 63 \, A^{-1/3}$ MeV = 17.3 MeV for $A=48$
and with the mean energy $\bar E$ = 16.9 MeV obtained for $^{48}$Ca
in \cite{KST}
in the 1p1h+1p1h$\otimes$phonon+GSC$_{phonon}$ approximation.
The present results for IS GQR in $^{48}$Ca are also in good
agreement with those obtained in Refs.~\cite{HSZ1,HSZ2} within
the self-consistent continuum RPA (in Ref.~\cite{HSZ1}
the maximum of $S(E)$ was obtained at 16.6 MeV, and
87~\% of the IS E2 EWSR was found within the 14--19 MeV interval).
However, in our calculation the peak energy of IS GQR
in $^{48}$Ni is about 1.8 MeV shifted upwards from the value
of 16.0 MeV obtained in Ref.~\cite{HSZ2}. This considerable
discrepancy seems to be explained by the difference between
the single-particle schemes and between the effective forces
used in the calculations being compared.
The case of the $^{48}$Ni nucleus appears to be more sensitive
to this difference due to the small proton separation energy.

From comparison of the results presented in Figs.~\ref{fig1}
and \ref{fig2} one can see that the proximity of the
proton separation energy to zero in $^{48,49}$Ni is especially
manifested in the case of $1 \hbar \omega$ IS E1 resonance
giving rise to a noticeable increase of its escape width
$\Gamma^{\uparrow}$ and to the larger depletion of the IS EWSR 
(about 35~\%) as compared with $^{48}$Ca
and $^{49}$Sc (about 22~\%). The broading of the IS E2
resonances in the unstable nuclei can be 
also seen from the results presented in Table~\ref{tab2}.

Note, however, that
our calculations cannot pretend to describe
total width of the resonances because only two main damping
mechanisms (Landau and escape ones) were taken into account
while the third one, i.~e. the spreading-width formation
mechanism, was not included in the calculations
(for such kind of calculations in $^{48}$Ca see Ref.~\cite{KST}).

\begin{figure}
\includegraphics*[scale=0.84]{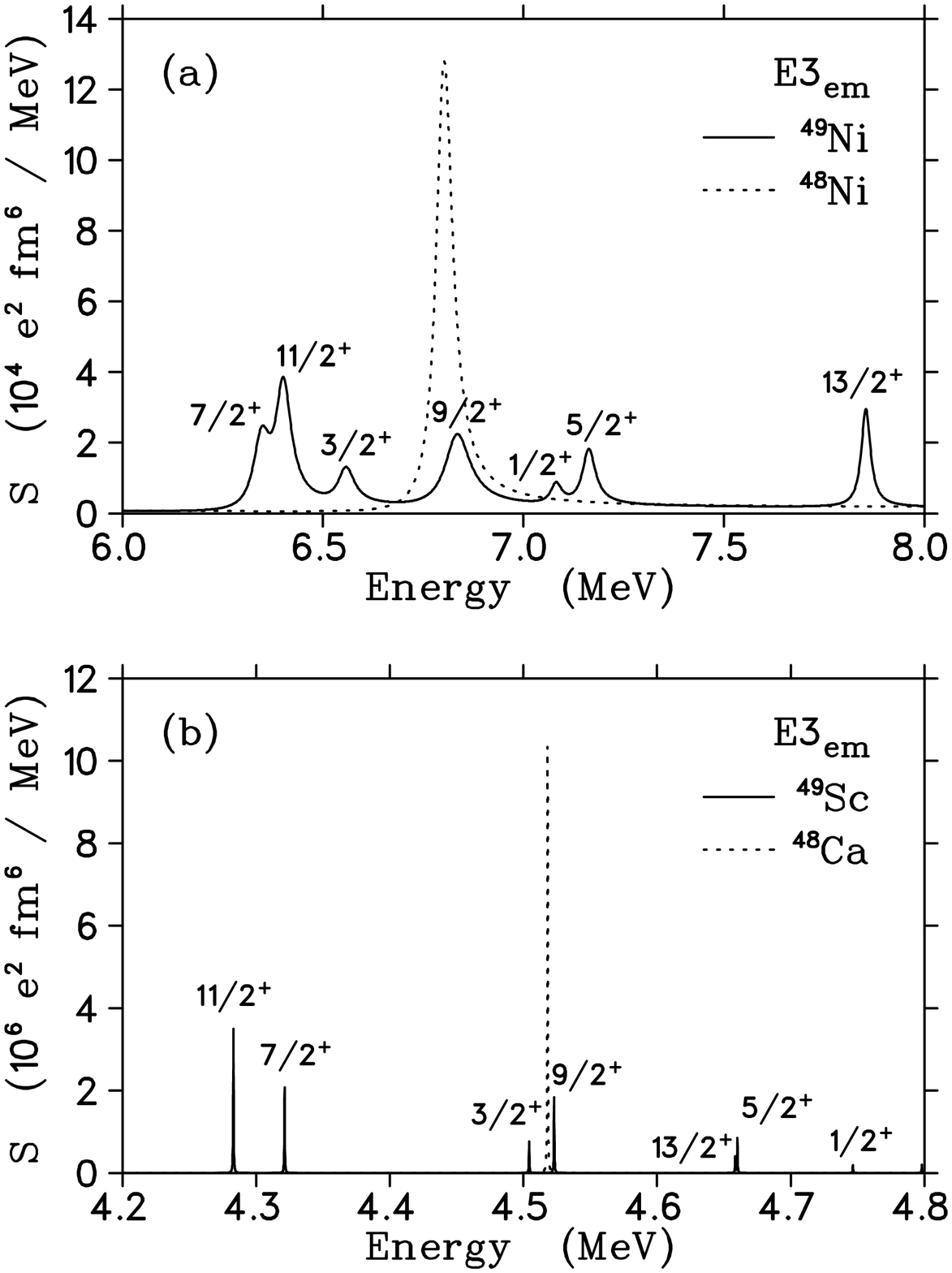}
\caption{\label{fig3}
The electromagnetic E3 strength functions $S(E)$
for $^{49}$Ni (a) and $^{49}$Sc (b) calculated
in the ORPA (solid lines), and the same functions for
$^{48}$Ni (a) and $^{48}$Ca (b)
calculated in the continuum TFFS (dotted lines).
The smearing parameter is $\Delta$ = 0.2 keV.}
\end{figure}
At last but not the least, in Fig.~\ref{fig3} we show the
strength functions for the $\{ f_{7/2} \otimes 3^- \}$
multiplets in $^{49}$Ni and $^{49}$Sc together
with the $3_1^-$ levels in $^{48}$Ni and $^{48}$Ca,
respectively. Here the neutron $1 f_{7/2}$
single-particle level corresponds to the ground state (g.s.)
of the $^{49}$Ni nucleus, the proton $1 f_{7/2}$ level
corresponds to the g.s. of the $^{49}$Sc.
As compared with the giant resonances  (Figs.~\ref{fig1}, \ref{fig2})
we see a very natural and noticeable role of the
odd neutron in $^{49}$Ni and proton in $^{49}$Sc.
All seven multiplet members are around the $3_1^-$ level
of the core at 6.8 MeV in the case of $^{49}$Ni and at 4.5 MeV
in the case of $^{49}$Sc. Since the $3_1^-$ level in $^{48}$Ni
lies in the continuum (and consequently is a resonance),
each of the multiplet members in $^{49}$Ni has a noticeable
escape width $\Gamma^{\uparrow}$.
Our approach enables one to calculate the escape width immediately
because the single-particle continuum is included completely.
The value of $\Gamma^{\uparrow}$ was determined
as a width at half maximum of the corresponding strength
function (note that the smearing parameter
$\Delta$ = 0.2 keV in Fig.~\ref{fig3}). The calculated value for
$^{48}$Ni is $\Gamma^{\uparrow} (3_1^-)$ = 51 keV.
For the multiplets members, the values of $\Gamma^{\uparrow}$,
energy shifts $\delta E$, and $B(E3,\, j^{\pi} \to g.s.)$
are listed in Table~\ref{tab3}.
\begin{table}
\begin{center}
\caption{\label{tab3}
Characteristics of the septuplet members
$\{ f_{7/2} \otimes 3^{-} \}_{j^{\pi}}$
in $^{49}$Ni and $^{49}$Sc. The calculated energy shifts
$\delta E$ from the excitation energies of $3^{-}_1$ states
in  $^{48}$Ni ($E(3^{-}_1)$ = 6.802 MeV)
and $^{48}$Ca ($E(3^{-}_1)$ = 4.518 MeV) are shown.
The escape widths $\Gamma^{\uparrow}$ for the septuplet
members in $^{49}$Ni are given. The values of
$B(E3,\, j^{\pi} \to g.s.)$ are shown in Weisskopf units
(1~W.~u. = 142.6 $e^2 fm^6$).}
\vspace{1em}
\begin{tabular}{rcrrrcrr}
\hline
\hline
\multicolumn{1}{c}{$j^{\pi}$} &
\multicolumn{1}{c}{} &
\multicolumn{3}{c}{$^{49}$Ni} &
\multicolumn{1}{c}{} &
\multicolumn{2}{c}{$^{49}$Sc}\\
\multicolumn{1}{c}{} &
\multicolumn{1}{c}{$\hphantom{abc}$} &
\multicolumn{1}{c}{$\delta E$} &
\multicolumn{1}{c}{$\Gamma^{\uparrow}$} &
\multicolumn{1}{c}{$B(E3)$} &
\multicolumn{1}{c}{$\hphantom{abc}$} &
\multicolumn{1}{c}{$\delta E$} &
\multicolumn{1}{c}{$B(E3)$} \\
\multicolumn{1}{c}{} &
\multicolumn{1}{c}{} &
\multicolumn{1}{c}{(keV)} &
\multicolumn{1}{c}{(keV)} &
\multicolumn{1}{c}{(W.~u.)} &
\multicolumn{1}{c}{} &
\multicolumn{1}{c}{(keV)} &
\multicolumn{1}{c}{(W.~u.)} \\
\hline
 ${\frac{ 1}{2}}^{+}$ &&  279 & 33 & 11.1 &&  228 &  3.4 \\
 ${\frac{ 3}{2}}^{+}$ && -244 & 49 & 13.9 &&  -14 &  6.7 \\
 ${\frac{ 5}{2}}^{+}$ &&  361 & 39 & 11.4 &&  142 &  5.1 \\
 ${\frac{ 7}{2}}^{+}$ && -455 & 52 & 14.7 && -197 & 10.1 \\
 ${\frac{ 9}{2}}^{+}$ &&   33 & 77 & 16.2 &&    5 &  6.9 \\
 ${\frac{11}{2}}^{+}$ && -401 & 46 & 14.3 && -235 & 10.3 \\
 ${\frac{13}{2}}^{+} \vphantom{\frac{1}{Q_Q}^{+}}$
                      && 1052 & 28 &  6.7 &&  140 &  1.0 \\
\hline
\hline
\end{tabular}
\end{center}
\end{table}
It is of great interest to measure the resonances in $^{49}$Ni 
and their "pure" escape widths, which may be an easier task
at present than the measurement of the $3_1^-$
level in $^{48}$Ni.

In conclusion, we have calculated some excitations in
$^{48}$Ni, $^{49}$Ni, $^{48}$Ca and $^{49}$Sc
within the same calculational scheme and with the consistent
account for the single-particle continuum. Despite
the fact that more complex configurations have not been
accounted for in the calculations,  our results expected to be
realistic enough to be compared with experiment.
In all four nuclei considered a distinct splitting of the IS E1
resonance into two peaks at about 11 MeV and 30 MeV has been
obtained. The influence of the small proton separation energy
in $^{48,49}$Ni has been demonstrated, especially on the
escape width and a large depletion of the IS E1 EWSR
by the $1 \hbar \omega$ IS E1 resonance.
For the $\{ f_{7/2} \otimes 3^- \}$
multiplet in $^{49}$Ni, the widths of all the multiplet
members are formed  by the single-particle
continuum in our approach.  Measurements of these excitations,
including escape widths of the individual states,
will give valuable information about many features 
of drip-line nuclei.

The authors are very grateful to Prof. V.~R.~Brown for 
her advices and careful reading of  the manuscript. 
\newpage
\end{document}